\documentclass[12pt]{article}
\pdfoutput=1

\usepackage{amsmath,amssymb}
\usepackage{graphicx}

\makeatletter

\renewcommand\section{\@startsection{section}{1}{\z@}
                                   {-3.5ex \@plus -1ex \@minus -.2ex}
                                   {2.3ex \@plus .2ex}
                                   {\normalfont\large\bfseries}}
\renewcommand\subsection{\@startsection{subsection}{2}{\z@}
                                   {-3.25ex\@plus -1ex \@minus -.2ex}
                                   {1.5ex \@plus .2ex}
                                   {\normalfont\normalsize\bfseries}}
\renewcommand\subsubsection{\@startsection{subsubsection}{3}{\z@}
                                   {-3.25ex\@plus -1ex \@minus -.2ex}
                                   {1.5ex \@plus .2ex}
                                   {\normalfont\normalsize\bfseries}}
\renewcommand\paragraph{\@startsection{paragraph}{4}{\z@}
                                   {3.25ex \@plus1ex \@minus.2ex}
                                   {-1em}
                                   {\normalfont\normalsize\bfseries}}

\makeatother

\newcommand{\beq}{\begin{equation}}
\newcommand{\eeq}{\end{equation}}
\newcommand{\bea}{\begin{eqnarray}}
\newcommand{\eea}{\end{eqnarray}}

\newcommand{\SL}{{\rm SL}}
\newcommand{\SU}{{\rm SU}}
\newcommand{\SO}{{\rm SO}}
\newcommand{\Sp}{{\rm Sp}}
\newcommand{\Spin}{\rm Spin}

\newcommand{\C}{\mathbb C}

\newcommand{\R}{\mathbb R}
\newcommand{\Z}{\mathbb Z}

\newcommand{\id}{\hbox{1\kern-.27em l}}

\newcommand{\Tr}{{\rm Tr}}

\newcommand{\cC}{{\cal C}}
\newcommand{\cD}{{\cal D}}
\newcommand{\cE}{{\cal E}}

\newcommand{\cH}{{\cal H}}

\newcommand{\cL}{{\cal L}}
\newcommand{\cM}{{\cal M}}

\newcommand{\cO}{{\cal O}}

\begin{document}

\pagestyle{empty}

\begin{center}

\vspace*{30mm}
{\LARGE Ground states of supersymmetric \\ Yang-Mills-Chern-Simons theory}

\vspace*{30mm}
{\large M{\aa}ns Henningson}

\vspace*{5mm}
Department of Fundamental Physics\\
Chalmers University of Technology\\
S-412 96 G\"oteborg, Sweden\\[3mm]
{\tt mans@chalmers.se}

\vspace*{30mm}{\bf Abstract:}
\end{center}
We consider minimally supersymmetric Yang-Mills theory with a Chern-Simons term on a flat spatial two-torus. The Witten index may be computed in the weak coupling limit, where the ground state wave-functions localize on the moduli space of flat gauge connections. We perform such computations by considering this moduli space as an orbifold of a certain flat complex torus. Our results agree with those obtained previously by instead considering the moduli space as a complex projective space. An advantage of the present method is that it allows for a more straightforward determination of the discrete electric 't~Hooft fluxes of the ground states in theories with non-simply connected gauge groups. A consistency check is provided by the invariance of the results under the mapping class group of a (Euclidean) three-torus. 

\newpage \pagestyle{plain}

\section{Introduction}
In three space-time dimensions, the usual Yang-Mills action for a gauge field $A$ with coupling constant $e$ 
\beq
S_\mathrm{YM} = \frac{1}{4 e^2} \int \Tr \left( F \wedge * F \right)
\eeq
may be complemented by a Chern-Simons term at level $k$ \cite{Deser-Jackiw-Templeton}
\beq
S_\mathrm{CS} = \frac{k}{4 \pi} \int \Tr \left(A \wedge d A + \frac{2}{3} A \wedge A \wedge A \right).
\eeq
For a connected gauge group $G$, invariance under `large' gauge transformations with a non-trivial space-time winding number valued in $\pi_3 (G) = \Z$ imposes a quantization condition on $k$: For a simply connected gauge group we must have $k = 0 \mod \Z$. For a gauge group of adjoint form, i.e. $G = \hat{G} / C$ where $\hat{G}$ is the simply connected covering group of $G$ and $C$ is the center subgroup of $\hat{G}$, the condition is $k = 0 \mod h \Z$ with $h$ denoting the dual Coxeter number of $G$.\footnote{This is possible because all fields transform in the adjoint representation of $\hat{G}$ and thus are invariant under $C$. Intermediate cases, in which $\hat{G}$ is divided by a proper subgroup of its center $C$, are also possible.}
\footnote{
For simplicity and clarity, we will be mostly concerned with the cases $\hat{G} = \SU (n)$ with center $C \simeq \Z / n \Z$, dual Coxeter number $h = n$, and $\Tr$ denoting the trace in the fundamental representation. These cases are in a certain sense the most interesting, but in principle we foresee no difficulties in generalizing to other groups.}

These models allow for a minimally supersymmetric extension by adding terms
\beq
S_\mathrm{Fermion} = \frac{1}{4 e^2} \int d^3 x \Tr \left( \bar{\lambda} D \!\!\!\! / \lambda \right) + \frac{k}{4 \pi} \int \Tr \left(\bar{\lambda}{\lambda} \right) ,
\eeq
where the fermionic field $\lambda$ is a Majorana spinor in the adjoint representation of $G$. Integrating out $\lambda$ gives back the bosonic model, but with the Chern-Simons level shifted \cite{Kao-Lee-Lee}\cite{Amelino-Camelia-Kogan-Szabo} to\footnote{
Here we assume that $k > 0$. See \cite{Witten99} for a discussion of negative levels.}
\beq
k^\prime = k - h / 2 .
\eeq
It is this shifted level $k^\prime$ that must obey the above quantization conditions, so in terms of the original level $k$ we must have
\beq
k = h / 2 \mod \Z \mathrm{\;\;\;\;\; or \;\;\;\;\;} k = h / 2 \mod h \Z
\eeq
for $G$ simply connected or of adjoint form respectively. 

The Riemannian structure together with the orientation induces a complex structure on two-dimensional space. In this paper, we will consider the case when space is a flat two-torus $T^2$, which we represent as
\beq
T^2 = \C / \Gamma
\eeq
for some rank two lattice $\Gamma \subset \C$. The size of $T^2$ can be absorbed in a rescaling of the coupling constant $e$ (of mass dimension $\frac{1}{2}$), so only the complex structure matters. We thus take $\Gamma$ to be generated by $1$ and a modular parameter $\tau$ in the complex upper half-plane. The supercharges $Q_+$ and $Q_-$, where the subscript indicates the chirality in two-dimensional space, are each others Hermitian conjugates and obey the supersymmetry algebra
\bea
\left\{ Q_+, Q_+ \right\} & = & P_{z} \cr
\left\{ Q_-, Q_- \right\} & = & P_{\bar{z}} \cr
\left\{ Q_+, Q_- \right\} & = & 2 H .
\eea
Here $P_{z}$ and $P_{\bar{z}}$ are the holomorphic and anti-holomorphic components of the spatial momentum, and $H$ is the Hamiltonian.

As usual, the $H = 0$ ground states are precisely the states annihilated by $Q_+$ and $Q_-$, and the graded dimension $I_{k^\prime}^G$ of the space of such states is independent of the coupling constant $e$ \cite{Witten82}. For a gauge group of adjoint form $G = \hat{G} / C$ there is a possible refinement: First of all, one may consider topologically non-trivial gauge bundles over the spatial $T^2$. These are classified by their discrete magnetic 't~Hooft flux \cite{'t_Hooft}
\bea
m_{12} & \in & H^2 (T^2, C) \cr
& \simeq & C .
\eea
We may also consider the behavior of the ground states under gauge transformations with non-trivial winding around the two cycles of $T^2$. Such winding is described by an element of $\mathrm{Hom} (\pi_1 (T^2), C)$, and the ground states may be classified by their transformation properties under this group, i.e. by a discrete electric 't~Hooft flux 
\bea
(e_1, e_2) & \in & \mathrm{Hom} (\mathrm{Hom} (\pi_1 (T^2), C), U (1)) \cr
& \simeq & H^1 (T^2, C) \cr
& \simeq & C \times C .
\eea
Here we have used the canonical isomorphism \cite{Witten00}
\bea
C & \simeq & C^* \cr
& \equiv & \mathrm{Hom} (C, U (1))  .
\eea
We let $I_{k^\prime}^G (e_1, e_2; m_{12})$ denote the graded dimension of the space of ground states with the indicated 't~Hooft fluxes. It is convenient to unify the electric and magnetic 't~Hooft fluxes by introducing the Fourier transform 
\beq
\hat{I}_{k^\prime}^G (m_{23}, m_{31}, m_{12}) = \sum_{(e_1, e_2) \in C \times C} I_{k^\prime} (e_1, e_2; m_{12}) \exp \left( 2 \pi i (m_{23} \cdot e_1 + m_{31} \cdot e_2) \right) ,
\eeq
where 
\bea
\left( m_{23}, m_{31}, m_{12} \right) & \in & H^2 (T^3, C) \cr
& \simeq & C \times C \times C .
\eea
Clearly
\bea \label{summed}
I_{k^\prime}^{\hat{G}} & = & \sum_{(e_1, e_2) \in C \times C} I_{k^\prime}^G (e_1, e_2; 0) \cr
& = & \hat{I}_{k^\prime}^G (0, 0, 0) .
\eea

In a Lagrangian formulation, $\hat{I}_{k^\prime}^G \left(m_{23}, m_{31}, m_{12} \right)$ can be interpreted as a path integral over field configurations with the appropriate 't~Hooft flux on a three-torus $T^3 = T^2 \times S^1$, where the second factor is the (compact, Euclidean) time direction:
\beq
\hat{I}_{k^\prime}^G \left(m_{23}, m_{31}, m_{12} \right) = \int \cD A \cD \lambda \, \exp (-S) .
\eeq
In this formulation, it is clear that the $\hat{I}_{k^\prime}^G \left(m_{23}, m_{31}, m_{12} \right)$ must be invariant under the $\SL_3 (\Z)$ mapping class group of $T^3$ acting on $\left( m_{23}, m_{31}, m_{12} \right)$ in the natural way. This provides an important consistency check on the results. As we will see, it is then important to take the bosonic or fermionic statistics of the different states correctly into account.

In a Hamiltonian formulation, $\hat{I}_{k^\prime}^G \left(m_{23}, m_{31}, m_{12} \right)$ can instead be interpreted as a trace over the Hilbert space $\cH (m_{12})$ in the sector with magnetic 't~Hooft flux $m_{12}$:
\beq
\hat{I}_{k^\prime}^G \left(m_{23}, m_{31}, m_{12} \right) = \Tr_{\cH (m_{12})} \left( (-1)^F e^{-\beta H} T (m_{23}, m_{31}) \right) ,
\eeq
where $F$ is the fermion number operator, and $T (m_{23}, m_{31})$ is the operator that twists with large gauge transformations parametrized by $m_{23}$ and $m_{31}$ in the $1$- and $2$-directions of $T^2$ respectively. In the weak coupling limit $e \rightarrow 0$, the magnetic contribution to the energy density means that the wave-functionals of such ground states localize on the finite-dimensional moduli space $\cM$ of gauge fields $A$ (modulo gauge transformations) for which the spatial components of the curvature $F$ vanishes.\footnote{
It is convenient to work in temporal gauge, where the time component $A_0$ of the gauge field is put to zero, and describe the bosonic part of the state by a wave-functional of the spatial components of $A$.} 

An elegant computation of the number of ground states $I_{k^\prime}^G$ for a simply connected gauge group\footnote{
Equivalently one can work with a gauge group $G = \hat{G} / C$ of adjoint form by taking $m_{12} = 0$ and summing the contributions for all values of $(e_1, e_2) \in C \times C$ as in (\ref{summed}).}
$G$ was described in \cite{Witten99}: This was based on the fact that the moduli space $\cM$ of flat connections, which inherits a complex structure from $T^2$, is isomorphic as a complex manifold to a weighted projective space \cite{Looijenga}\cite{Bernshtein-Shvartsman}\cite{Friedman-Morgan-Witten}:
\beq
\cM \simeq \mathbb C \mathbb P^r_{s_0, s_1, \ldots, s_r} ,
\eeq
where the weights $s_0 = 1$ and $s_1, \ldots, s_r$ are given by the Dynkin coefficients of the highest coroot of $G$. The space of ground states can be identified with the cohomology group $H^0 (\cM, \cO ({k^\prime}))$.\footnote{
Here $\cO (k^\prime)$ is the $k^\prime$ power of the tautological line bundle $\cO (1)$ over the projective space $\cM$. The vanishing of the higher cohomology groups $H^i (\cM, \cO ({k^\prime}))$ for $i > 0$ is important.} This can be further identified with the space of homogeneous polynomials of weighted degree $k^\prime$ in the homogenous coordinates of $\cM$, and the dimension is thus readily computed. E.g. in the case of $\hat{G} = \SU (n)$ we have $r = n - 1$ and $s_1 = \ldots = s_r = 1$, so the moduli space is then an ordinary projective space $\cM = \mathbb C \mathbb P^{n - 1}$ and the dimension of the space of ground states is 
\beq \label{SU(n)groundstates}
I_{k^\prime}^{\SU (n)} = \frac{1}{(n - 1)!} (k^\prime + n - 1) \ldots (k^\prime + 1) .
\eeq
In particular, for $k^\prime = 0$ there is a unique ground state, and in \cite{Witten00} it was argued that this (when interpreted in the theory with gauge group $G = \hat{G} / C = \SU (n) / \Z_n$ of adjoint type) has trivial electric 't~Hooft flux $(e_1, e_2) = (0, 0)$ (and of course trivial magnetic 't~Hooft flux $m_{12} = 0$).  

In this paper, we will reanalyze this problem in a more pedestrian way by instead regarding the moduli space as a discrete quotient of the complex torus 
\beq
X = (\C \otimes V) / (\Gamma \otimes \Lambda)
\eeq
by the Weyl group $W$ of $G$, i.e.
\beq
\cM \simeq X / W .
\eeq 
Here $V$ is the root space and $\Lambda \subset V$ is the root lattice of $G$, so
\beq \label{maximal_torus}
U = V / \Lambda
\eeq
is a maximal torus of $G$. In the next section, we will define a certain holomorphic line-bundle $\cL^{k^\prime}$ over $X$ and describe the corresponding space of holomorphic sections $H^0 (X, \cL^{k^\prime})$. The Weyl group $W$ acts on this space, and the space of ground states is given by the $W$-invariant subspace $H^\mathrm{inv} \subset H^0 (X, \cL^{k^\prime})$.  

It is not immediately obvious that this method reproduces e.g. the result (\ref{SU(n)groundstates}), but we confirm it explicitly in section three for the cases with gauge group $\SU (2)$ or $\SU (3)$. We also determine the spectrum of 't~Hooft fluxes in the theories with gauge group $\SU (2) / \Z_2$ and $\SU (3) / \Z_3$ when $k^\prime = 0 \mod 2$ and $k^\prime = 0 \mod 3$ respectively. 

E.g the spectrum of 't~Hooft fluxes seems more difficult to obtain in a computation based on regarding $\cM$ as a projective space than in the present approach where $\cM$ is regarded as an orbifold. Hopefully, this viewpoint will be useful in future studies of these models. In particular, I plan to use it to study the spectrum of positive energy states, to begin with in the weak coupling limit (or equivalently small torus) limit, but later also in perturbation theory.

I recently became aware of the references \cite{Smilga}, which have some overlap with the present work.

\section{The spectrum of ground states}
A flat connection $A$ on a principal $G$-bundle over $T^2$ is determined by its two commuting holonomies around a basis of one-cycles. By a gauge transformation, which acts by simultaneous conjugation by an element of $G$, we may take the holonomies to be elements of the maximal torus (\ref{maximal_torus}). The remaining gauge transformations amount to the Weyl group $W$ of $G$ acting by linear transformations on $V$ preserving $\Lambda$. 

The two holonomies may be combined to an element $Z$ of the complex torus
\beq
X = \C \otimes V / (\Gamma \otimes \Lambda) .
\eeq
We can think of $Z$ as a constant gauge field over $T^2$ with values in $V$, and this is the only part of $A$ that is relevant at low energies. Similarly, we need only consider the spatially constant $V$-valued modes $\eta_+$ and $\eta_-$ of the spinor field $\lambda$.\footnote{Along certain divisors of $X$, some non-abelian gauge symmetry remains unbroken and there are further low-energy degrees of freedom. The dynamics of these is described by the dimensionally reduced theory, i.e. by supersymmetric quantum mechanics with three supercharges. However, it is believed that the latter system does not have any bound state at threshold \cite{Hoppe-Yau}, so this symmetry restoration will not affect our counting of ground states.} The low-energy action thus reads
\bea
S & = & \frac{1}{4 e^2} \int d t \left( \frac{d Z}{d t} \cdot \frac{d \bar{Z}}{d t} + \eta_+ \cdot \frac{d \eta_-}{d t} + \eta_- \cdot \frac{d \eta_+}{d t} \right) \cr
& & + \frac{\pi k^\prime}{\tau - \bar{\tau}} \int d t \left(  Z \cdot \frac{d \bar{Z}}{d t} - \bar{Z} \cdot \frac{d Z}{d t} + \eta_+ \cdot \eta_- \right) ,
\eea  
where the raised dot denotes evaluation of the Cartan matrix\footnote{
This is the restriction to $V$ of the invariant form $\Tr$ on the Lie algebra of $G$.}
$\cC$ of $G$ on two $V$-valued quantities. Note that the Chern-Simons level has been shifted from $k$ to $k^\prime$ as described in the introduction (although we have only integrated out the massive fermionic modes). The low-energy supercharges are 
\bea
Q_+ & = & \frac{1}{4 e^2} \eta_- \cdot \frac{d Z}{d t}  \cr
Q_- & = & \frac{1}{4 e^2} \eta_+ \cdot \frac{d \bar{Z}}{d t} .
\eea

To understand the action of $Q_+$ and $Q_-$ on the Hilbert space of the theory, we begin by determining the momenta $\bar{\Pi}$ and $\Pi$ conjugate to $Z$ and $\bar{Z}$:
\bea
\bar{\Pi} & = & \frac{1}{4 e^2} \frac{d \bar{Z}}{d t} - \frac{\pi k^\prime}{\tau - \bar{\tau}} \bar{Z} \cr
\Pi & = & \frac{1}{4 e^2} \frac{d Z}{d t} + \frac{\pi k^\prime}{\tau - \bar{\tau}} Z .
\eea
These momenta act on a wave-function $\Psi (Z, \bar{Z})$ as $i \frac{\partial}{\partial Z}$ and $i \frac{\partial}{\partial \bar{Z}}$ respectively, so the supercharges can be rewritten as
\bea
Q_+ & = & i \eta_- \cdot \frac{D}{D Z} \cr
Q_- & = & i \eta_+ \cdot \frac{D}{D \bar{Z}} ,
\eea
where the holomorphic and anti-holomorphic covariant derivatives are given by
\bea
\frac{D}{D Z} & = & \frac{\partial}{\partial Z} - \frac{i \pi k^\prime}{\tau - \bar{\tau}} \bar{Z} \cr
\frac{D}{D \bar{Z}} & = & \frac{\partial}{\partial \bar{Z}} + \frac{i \pi k^\prime}{\tau - \bar{\tau}} Z .
\eea
Since the holomorphic covariant derivatives commute with each other, they define a holomorphic line-bundle $\cL^{k^\prime}$ over the complex torus $X$. The curvature of this line bundle is constant over $X$ and given by
\beq
\left[ \frac{D}{D Z}, \frac{D}{D \bar{Z}} \right] = k^\prime \omega ,
\eeq
with 
\beq
\omega = \frac{1}{\tau_2} d Z \cdot \wedge d \bar{Z} .
\eeq
The normalization of $\omega$ is such that it defines an element of $H^2 (X, \Z)$, whereas all smaller multiples of it would only give a real cohomology class. For a more extensive description of such holomorphic line bundles over complex tori, see e.g. \cite{Griffiths-Harris}.

The canonical anti-commutation relations of the fermionic degrees of freedom define a Clifford algebra
\bea
\left\{ \eta_+, \eta_+ \right\} & = & 0 \cr
\left\{ \eta_-, \eta_- \right\} & = & 0 \cr
\left\{ \eta_+, \eta_- \right\} & = & \cC
\eea
with $\cC$ again being the Cartan matrix of $G$. This can be represented on a $2^r$-dimensional space, constructed by acting with the components of $\eta_+$ on a Weyl invariant state $| \Omega_- \rangle$ annihilated by all components of $\eta_-$. In fact, the ground states of the theory will be of the form
\beq
\Psi (Z, \bar{Z}) \otimes | \Omega_- \rangle
\eeq
with some wave-function $\Psi (Z, \bar{Z})$. Such a state is automatically annihilated by $Q_+$, and the requirement that it be annihilated also by $Q_-$ amounts to the holomorphicity condition
\beq \label{holomorphicity}
 \frac{D}{D \bar{Z}} \Psi (Z, \bar{Z}) = 0 .
 \eeq
The Riemann-Roch theorem gives the dimension of the space of such holomorphic sections of the line bundle $\cL^{k^\prime}$ over the flat manifold $X$ as
\bea
\dim H^0 (X, \cL^{k^\prime}) & = & \int_X e^{k^\prime \omega}  \cr
& = & \frac{(k^\prime)^r}{r !} \int_X \omega^r \cr
& = & (k^\prime)^r c ,
\eea
where $r$ is the rank of $G$ and $c$ is the determinant of the Cartan matrix. We have
\bea
c & = & \det \cC \cr
& = & \left| \Lambda^* / \Lambda \right| \cr
& = & \left| C \right|
\eea
where the weight lattice $\Lambda^*$ is the dual of the root lattice $\Lambda$ of $G$.\footnote{
For simplicity, we restrict attention to the case of simply laced gauge groups i.e. $\hat{G} = \SU (n), \Spin (2 r), E_6, E_7$, or $E_8$. The generalization to the non-simply laced groups $\hat{G} = \Spin (2 r + 1), \Sp (2 r), F_4$, or $G_2$ should not entail any conceptual difficulties, though, especially since their center subgroups contain at most one non-trivial element and their Weyl groups are well understood.} 
In the last line we have used that the quotient group $\Lambda^* / \Lambda$ is isomorphic to the center subgroup $C$ of $G$. For the case of $G = \SU (n)$ we have $c = n$.

It is convenient to change to a holomorphic trivialization of $\cL^{k^\prime}$ by writing
\beq
\Psi (Z, \bar{Z}) = \exp \left( \frac{i \pi k^\prime}{\tau - \bar{\tau}} (Z - \bar{Z}) \cdot Z \right) \psi (Z, \bar{Z}) .
\eeq
In terms of $\psi$, the holomorphicity condition (\ref{holomorphicity}) simply amounts to 
\beq
\frac{\partial}{\partial \bar{Z}} \psi = 0 .
\eeq
To obtain an explicit description of the space of holomorphic sections $H^0 (X, \cL^{k^\prime})$, we consider the pullback by the projection map
\beq
\pi \colon \C \otimes V \rightarrow X .
\eeq
This takes an element of $H^0 (X, \cL^{k^\prime})$ to a holomorphic function $\psi$ on $\C \otimes V$ that obeys the quasi-periodicity conditions
\beq
\psi (Z + \alpha + \tau \beta) = \exp \left( -i \pi k^\prime \tau \beta \cdot \beta - 2 \pi i k^\prime \beta \cdot Z \right) \psi (Z)
\eeq
for arbitrary elements $\alpha$ and $\beta$ of the root lattice $\Lambda$. The Chern class $k^\prime [\omega]$ is reflected in the second term in the exponential prefactor. In this trivialization, the transition functions are thus holomorphic. In terms of $\Psi$, the transformation law reads
\bea
\Psi (Z {+} \alpha {+} \tau \beta, \bar{Z} {+} \alpha {+} \bar{\tau} \beta) & = & \exp \left( i \pi k^\prime \left(\frac{\alpha \cdot (Z {-} \bar{Z}) + \bar{\tau} \beta \cdot Z {-} \tau \beta \cdot \bar{Z}}{\tau - \bar{\tau}} + \alpha \cdot \beta \right) \right) \cr
& & \times \Psi (Z, \bar{Z}) ,
\eea
so in the original trivialization the transition functions are instead $U (1)$-valued.

However, since $X$ is non-simply connected, a holomorphic line bundle is determined by its curvature only up to translation (by an element of the torus $X$ itself via its abelian group structure). What makes $\cL^{k^\prime}$ distinguished among its translates is its symmetry properties: The above transformation law is clearly covariant under the Weyl group $W$ (acting on the roots $\alpha$, $\beta$ and the variable $Z$ in the natural way). It is also covariant under the $\SL_2 (\Z)$ mapping class group of $T^2$.\footnote{
This may be compared with e.g. the transformation law $\theta (\tau | z + a + b \tau) = \exp ( - i \pi \tau b^2 - 2 \pi i b z) \theta (\tau | z)$ for $a, b \in \Z$ of the Jacobi theta function $\theta (\tau | z)$, which is {\it not} invariant under $\SL_2 (\Z)$. The difference is that the $\beta \cdot \beta$ is an {\it even} integer for all roots $\beta$. Indeed, the square of the Jacobi theta function is a section of an $\SL_2 (\Z)$ invariant line bundle over $T^2$.
}
Indeed, $\SL_2 (\Z)$ is generated by the two transformations
\bea
S \colon & & \tau \mapsto - 1 / \tau \cr
T \colon & & \tau \mapsto \tau + 1 .
\eea
Given a $\psi$ transforming as above, we define functions $\psi_S$ and $\psi_T$ with the same transformation properties by
\bea
\psi_S (\tau | Z) & = & \exp \left( - \frac{i \pi k^\prime}{\tau} Z \cdot Z \right) \psi ( - 1 / \tau | Z / \tau) \cr
\psi_T (\tau | Z) & = & \psi (\tau + 1 | Z) ,
\eea
where we have explicitly indicated also the dependence on $\tau$. 

Another important symmetry property of the line bundle $\cL^{k^\prime}$ is that it is invariant under translations by $\delta + \epsilon \tau$ for arbitrary elements $\delta, \epsilon \in \frac{1}{k^\prime} \Lambda^* / \Lambda$. Indeed, given a function $\psi$ transforming as above, we may define a translated function $T_{\delta + \epsilon \tau} \psi$ by
\beq \label{translations}
\left( T_{\delta + \epsilon \tau} \psi \right) (Z) = \pm \exp \left( i \pi k^\prime \epsilon \cdot (\delta + \epsilon \tau ) + 2 \pi i k^\prime \epsilon \cdot Z \right) \psi (Z + \delta + \epsilon \tau)
\eeq
with the same transformation properties. This is well-defined only up to the ambigious sign: Shifting $\delta \rightarrow \delta + \alpha$ and $\epsilon \rightarrow \epsilon + \beta$ with $\alpha, \beta \in \Lambda$ multiplies the right hand side by the sign $\exp \left( i \pi k^\prime (\alpha \cdot \epsilon - \beta \cdot \delta) \right)$. Also, it should be noted that these translation operators in general do not commute, but only fulfill the Heisenberg algebra
\beq \label{T-algebra}
T_{\delta + \epsilon \tau} T_{\delta^\prime + \epsilon^\prime \tau}  = \pm \exp \left( i \pi k^\prime (\epsilon^\prime \cdot \delta - \epsilon \cdot \delta^\prime) \right) T_{\delta + \delta^\prime + (\epsilon + \epsilon^\prime) \tau} .
\eeq
The translations operators $T_{\epsilon \tau}$ (i.e. with $\delta = 0$) are convenient for constructing a basis of $H^0 (X, \cL^{k^\prime})$. Let $\psi \in H^0 (X, \cL^{k^\prime})$ be given by
\beq
\psi (Z) = \sum_{\lambda \in \Lambda} \exp \left(i \pi k^\prime \tau \lambda \cdot \lambda + 2 \pi i k^\prime \lambda \cdot Z \right) . 
\eeq
A basis for $H^0 (X, \cL^{k^\prime})$ is then given by the translated functions $T_{\epsilon \tau} \psi$ for $\epsilon \in  \frac{1}{k^\prime} \Lambda^* / \Lambda$:
\beq
(T_{\epsilon \tau} \psi) (Z) = \sum_{\lambda \in \Lambda} \exp \left( i \pi k^\prime \tau (\lambda + \epsilon) \cdot (\lambda + \epsilon) + 2 \pi i k^\prime (\lambda + \epsilon) \cdot Z \right) .
\eeq

The Weyl group $W$ acts on the basis elements $T_{\epsilon \tau} \psi$ by permutation, and $ \frac{1}{k^\prime} \Lambda^* / \Lambda$ is thus partitioned into a disjoint union of Weyl orbits $\cE_s$:
\beq
\frac{1}{k^\prime} \Lambda^* / \Lambda  = \bigcup_{s \in S} \cE_s 
\eeq
The dimension of the Weyl invariant subspace of $H^\mathrm{inv} \subset H^0 (X, \cL^{k^\prime})$ equals the cardinality of the set $S$ of orbits. Indeed, for each $s \in S$ we can construct a Weyl invariant function $\Xi_s \in H^0 (X, \cL^{k^\prime})$  as 
\beq
\Xi_s = \sum_{\epsilon \in \cE_s} T_{\epsilon \tau} \psi .
\eeq
But it is not quite obvious that this dimension agrees with the results described in the introduction, e.g. the formula (\ref{SU(n)groundstates}) in the case of $\hat{G} = \SU (n)$. In the next section, we will verify this for $\hat{G} = \SU (2)$ and $\hat{G} = \SU (3)$, but a more general understanding would be desirable.

In the case that $k^\prime = 0 \mod h \Z$, so that the theory with an adjoint type gauge group is defined, the smaller set of translation operators $T_{\mu + \nu \tau}$ for $\mu, \nu \in \Lambda^* /  \Lambda$ are of particular interest. There is then a canonical choice of signs in (\ref{translations}):
\beq
\left( T_{\mu + \nu \tau} \psi \right) (Z) = \exp \left( i \pi k^\prime \mu \cdot \mu + i \pi k^\prime \nu \cdot \nu + i \pi k^\prime  \tau \nu \cdot \nu + 2 \pi i k^\prime \nu \cdot Z \right) \psi (z + \mu + \nu \tau) .
\eeq
Shifting $\mu \rightarrow \mu + \alpha$ and $\nu \rightarrow \nu + \beta$ with $\alpha, \beta \in \Lambda$ now leaves the right hand side invariant, so the definition is unambigious. The prefactor may seem ad hoc, but is in fact determined by $\SL_2 (\Z)$ covariance as will become apparent later. These operators obey the commuting algebra
\beq
T_{\mu + \nu \tau} T_{\mu^\prime + \nu^\prime \tau} = T_{\mu + \mu^\prime + (\nu + \nu^\prime) \tau} .
\eeq
The transformation properties of a physical state under this commuting translations group determine the electric 't~Hooft flux $(e_1, e_2) \in C \times C$.  Indeed, acting with the translation operator $T_{\mu + \nu \tau}$ on the basis element $T_{\epsilon^\prime \tau} \psi$ gives another basis element multiplied by a phase factor:
\beq
\left( T_{\mu + \nu \tau} T_{\epsilon \tau} \psi \right) (\psi) = \exp \left( i \pi k^\prime \mu \cdot \mu + i \pi k^\prime \nu \cdot \nu + 2 \pi i k^\prime \epsilon \cdot \mu \right) \left( T_{(\nu + \epsilon) \tau} \psi \right) (Z) .
\eeq
The phase factor determines $e_1$, and states of definite $e_2$ are obtained by taking appropriate linear combinations of the states $T_{(\nu + \epsilon) \tau} \psi$ for $\nu \in \Lambda^* / \Lambda$.

\section{The case of $\hat{G} = \SU (n)$}
For any simply connected gauge group $\hat{G}$, the case $k^\prime = 0$ is somewhat exceptional: The line bundle is then trivial and there is a single section given by a constant function, which clearly is Weyl invariant. The corresponding state (in the theory with gauge group $G = \hat{G} / C$ of adjoint type) has trivial electric and magnetic 't~Hooft flux $(e_1, e_2) = (0, 0)$ (and of course $m_{12} = 0$). 

For $k^\prime = 1$, the $c = | \Lambda^* / \Lambda |$ different functions $T_{\epsilon \tau} \psi$ for $\epsilon \in \Lambda^* / \Lambda$ are all Weyl invariant. (The theory with adjoint type gauge group is then not defined, so there is no question about electric or magnetic 't~Hooft fluxes here.) 

For $k^\prime \geq 2$ things get more interesting. We will content ourselves with discussing the cases when $\hat{G} = \SU (n)$. The root space can then be identified with 
\beq
V = \left\{ (x_1, \ldots, x_n) \in \R^n | x_1 + \ldots + x_n = 0 \right\} 
\eeq
with the standard inner product on $\R^n$. The root lattice is
\beq
\Lambda = \left\{ (x_1, \ldots, x_n) \in \Z^n | x_1 + \ldots + x_n = 0 \right\} ,
\eeq
and the weight lattice is
\beq
\Lambda^* = \left\{ (x_1, \ldots, x_n) \in \left(\frac{1}{n} \Z \right)^n \Big| x_1 + \ldots + x_n = 0, \;\;\; x_1 = \ldots = x_n \mod \Z \right\} . 
\eeq
The Weyl group is given by the group of permutations of the entries of a vector $(x_1, \ldots, x_n) \in V$. 

The center subgroup of $\SU (n)$ is
\beq
C = \Lambda^* / \Lambda \simeq \Z / n \Z ,
\eeq
and is generated by the element 
\beq
\left( \frac{1}{n}, \ldots, \frac{1}{n}, - \frac{n - 1}{n} \right)
\eeq for which 
\beq
\left( \frac{1}{n}, \ldots, \frac{1}{n}, - \frac{n - 1}{n} \right) \cdot \left( \frac{1}{n}, \ldots, \frac{1}{n}, - \frac{n - 1}{n} \right) = \frac{n - 1}{n} .
\eeq

The dual Coxeter number of $\SU (n)$ is $h = n$, so if $k^\prime = 0 \mod n$, we may consider the theory with gauge group $G = \SU (n) / \Z_n$ . The states can then be classified by their magnetic 't~Hooft flux 
\bea
m_{12} & \in & C \cr
& \simeq & \Z / n \Z 
\eea
and their electric 't~Hooft flux 
\bea
(e_1, e_2) & \in & C \times C \cr
& \simeq & (\Z / n \Z) \times (\Z / n \Z) .
\eea
(This means that the eigenvalues under the translations 
\beq
T_{\left( \frac{1}{n}, \ldots, \frac{1}{n}, - \frac{n - 1}{n} \right)} \mathrm{\;\;\;\; and \;\;\;\;} T_{\left( \frac{1}{n}, \ldots, \frac{1}{n}, - \frac{n - 1}{n} \right) \tau}
\eeq 
are $\exp (2 \pi i e_1 / n)$ and $\exp (2 \pi i e_2 / n)$ respectively.) Not all combinations of fluxes are possible, though: Writing $n = u v$ with
\bea
u & = & \mathrm{GCD} (m_{12}, n) \cr
v & = & n / u ,
\eea
the holonomies break the gauge group to a subgroup of a group isomorphic to $\SU (u)$. Indeed, a flat connection on a bundle with magnetic 't~Hooft flux $m_{12} \in C \simeq \Z / n \Z$ is specified by two commuting holonomies $U_1$ and $U_2$ whose lifts $\hat{U}_1$ and $\hat{U}_2$ to $\SU (n)$ obey the almost commutation relations
\beq
\hat{U}_1 \hat{U}_2 = e^{2 \pi i m_{12} / n} \hat{U}_2 \hat{U}_1 .
\eeq
We can construct such holonomies as
\bea
\hat{U}_1 & = & \left(
\begin{matrix}
e^{i \pi m_{12} (-v + 1) / v} & 0 & \ldots & 0 \cr
0 & e^{i \pi m_{12} (-v + 3) / v} & \ddots & \vdots \cr
\vdots & \ldots & \ddots & \vdots \cr
0 & \ldots & 0 & e^{i \pi m_{12} (v - 1) / v} 
\end{matrix}
\right) \otimes u_1 \cr
\hat{U}_2 & = & \left(
\begin{matrix}
0 & 1 & \ldots & 0 \cr
\vdots & 0 & \ddots & \vdots \cr
0 & \ldots & 0 & 1 \cr
1 & 0 & \ldots & 0
\end{matrix}
\right) \otimes u_2 ,
\eea
where $u_1$ and $u_2$ are two arbitrary commuting $\SU (u)$ matrices. 

The electric 't~Hooft flux obeys
\bea
(e_1, e_1) & \in & (v \Z / n \Z) \times (v \Z / n \Z) \cr
& \simeq & (\Z / u \Z) \times (\Z / u \Z) .
\eea
We are here effectively considering an $\SU (u)$ gauge theory at the rescaled (and shifted) level $k^\prime / v$. The electric 't~Hooft fluxes with values in $\Z / u \Z$ should be rescaled by a factor of $v$ to be interpreted in the $\SU (n)$ theory as elements of $\Z / n \Z$. One must also determine the statistics (bosonic or fermionic) of these states. (By convention, we take the ground states with trivial magnetic 't~Hooft flux $m_{12} = 0$ to be bosonic.)

After Fourier transformation, the $\SL_3 (\Z)$ mapping class group of $T^3$ acts in the natural way on $(m_{23}, m_{31}, m_{12}) \in C \times C \times C$ via its reduction modulo $n$ so that the only invariant is
\beq
\bar{m} = \mathrm{GCD} (m_{23}, m_{31}, m_{12}, n) .
\eeq
Thus $\hat{I}_{k^\prime}^{\SU (n) / Z_n} (m_{23}, m_{31}, m_{12})$ must be a function of $\bar{m}$ only.

\subsection{$\hat{G} = \SU (2)$}
The elements of $\frac{1}{k^\prime} \Lambda^* / \Lambda$ can be represented by the $2 k^\prime$ different $\frac{1}{k^\prime} \Lambda^*$ elements
\beq
\left\{ \Bigl( 0, 0 \Bigr), \left(\frac{1}{2 k^\prime}, - \frac{1}{2 k^\prime} \right), \ldots, \left(\frac{2 k^\prime - 1}{2 k^\prime}, - \frac{2 k^\prime - 1}{2 k^\prime} \right) \right\} .
\eeq
The single non-trivial element of the Weyl group acts on these as
\beq
\left( \frac{l}{2 k^\prime}, - \frac{l}{2 k^\prime} \right) \mapsto \left( \frac{2 k^\prime - l}{2 k^\prime}, - \frac{2 k^\prime - l}{2 k^\prime} \right) .
\eeq
As shown in table \ref{SU(2)table}, there are thus 
\beq
I^{\SU (2)}_{k^\prime} = k^\prime + 1
\eeq
Weyl invariant ground states, in agreement with (\ref{SU(n)groundstates}).
\begin{table}[ht]
\centering
\begin{tabular}{r}
\\
state \cr
\hline \cr
$\psi$ \cr
\cr
$T_{\left(\frac{k^\prime}{2 k^\prime}, - \frac{k^\prime}{2 k^\prime} \right) \tau} \psi$ \cr
\cr
$\frac{1}{\sqrt{2}} \left( T_{\left(\frac{l}{2 k^\prime}, - \frac{l}{2 k^\prime} \right) \tau} + T_{\left(\frac{2 k^\prime - l}{2 k^\prime}, - \frac{2 k^\prime - l}{2 k^\prime} \right) \tau} \right) \psi$ \cr
\cr \hline \cr
\end{tabular}
\caption{States in the $\SU (2)$ theory for arbitrary $k^\prime$. In the last line $l = 1, \ldots, k^\prime - 1$.}
\label{SU(2)table}
\end{table}

If $k^\prime$ is even, we may consider the theory with gauge group $G = \SU (2) / \Z_2 \simeq \SO (3)$. The above states are then in the sector with trivial magnetic 't~Hooft flux $m_{12} = 0$, and may be further characterized by their electric 't~Hooft fluxes, i.e. by their eigenvalues $e_1$ and $e_2$ under the translations $T_{\left(\frac{1}{2}, - \frac{1}{2} \right)}$ and $T_{\left(\frac{1}{2}, - \frac{1}{2} \right) \tau}$. 

For $k^\prime = 0 \mod 4$ these translation eigenstates are shown in table \ref{k=0mod4table}. 
\begin{table}[ht] 
\centering
\begin{tabular}{rcc}
\\
state & $e_1$ & $e_2$ \cr
\hline \cr
$\frac{1}{\sqrt{2}} \left( T_{(0, 0) \tau} + T_{\left( \frac{1}{2}, - \frac{1}{2} \right) \tau} \right) \psi$ & $0$ & $0$ \cr
$\frac{1}{\sqrt{2}} \left( T_{(0, 0) \tau} - T_{\left( \frac{1}{2}, - \frac{1}{2} \right) \tau} \right) \psi$ & $0$ & $1$ \cr
\cr
$\frac{1}{\sqrt{2}} \left( T_{\left( \frac{1}{4},  - \frac{1}{4} \right) \tau} + T_{\left( \frac{3}{4}, - \frac{3}{4} \right) \tau} \right) \psi$ & $0$ & $0$ \cr
\cr
$\frac{1}{2} \left( T_{\left( \frac{l}{2 k^\prime}, -  \frac{l}{2 k^\prime} \right) \tau} + T_{\left( \frac{2 k^\prime - l}{2 k^\prime}, -  \frac{2 k^\prime - l}{2 k^\prime} \right) \tau} + T_{\left( \frac{k^\prime + l}{2 k^\prime}, -  \frac{k^\prime + l}{2 k^\prime} \right) \tau} +  T_{\left( \frac{k^\prime - l}{2 k^\prime}, -  \frac{k^\prime - l}{2 k^\prime} \right) \tau} \right) \psi$ & $l \mod 2$ & $0$ \cr
$\frac{1}{2} \left( T_{\left( \frac{l}{2 k^\prime}, -  \frac{l}{2 k^\prime} \right) \tau} + T_{\left( \frac{2 k^\prime - l}{2 k^\prime}, -  \frac{2 k^\prime - l}{2 k^\prime} \right) \tau} - T_{\left( \frac{k^\prime + l}{2 k^\prime}, -  \frac{k^\prime + l}{2 k^\prime} \right) \tau} -  T_{\left( \frac{k^\prime - l}{2 k^\prime}, -  \frac{k^\prime - l}{2 k^\prime} \right) \tau} \right) \psi$ & $l \mod 2$ & $1$ \cr 
\cr \hline \cr
\end{tabular}
\caption{States in the $\SU (2) / \Z_2$ theory for $k^\prime = 0 \mod 4$ and trivial magnetic 't~Hooft flux $m_{12} = 0$. In the last two lines $1 \leq l \leq k^\prime / 2 - 1$.}
\label{k=0mod4table}
\end{table}
There are thus $k^\prime / 4 + 1$ states that are singlets under $\SL_2 (\Z)$ with trivial electric 't~Hooft flux $(e_1, e_2) = (0, 0)$, and $k^\prime / 4$ triplets of states with non-trivial electric 't~Hooft fluxes $(e_1, e_2) = (0, 1), (1, 0), (1, 1)$:
\bea
I^{\SO (3)}_{k^\prime} (0, 0; 0) & = & \frac{k^\prime}{4} + 1 \cr
I^{\SO (3)}_{k^\prime} (e_1, e_2; 0) & = & \frac{k^\prime}{4} \mathrm{\;\;\; for \;\;\;} (e_1, e_2) \neq (0, 0) 
\eea
which after Fourier transformation gives
\bea \label{kprime=0mod4}
\hat{I}^{\SO (3)}_{k^\prime} (0, 0, 0) & = & k^\prime + 1 \cr
\hat{I}^{\SO (3)}_{k^\prime} (m_{23}, m_{31}, 0) & = & 1 \mathrm{\;\;\; for \;\;\;} (m_{23}, m_{31}) \neq (0, 0) . 
\eea

For $k^\prime = 2 \mod 4$ the translation eigenstates are shown in table \ref{k=2mod4table}. 
\begin{table}[ht] 
\centering
\begin{tabular}{rcc}
\\
state & $e_1$ & $e_2$ \cr
\hline \cr
$\frac{1}{\sqrt{2}} \left( T_{(0, 0) \tau} + T_{\left( \frac{1}{2}, - \frac{1}{2} \right) \tau} \right) \psi$ & $1$ & $1$ \cr
$\frac{1}{\sqrt{2}} \left( T_{(0, 0) \tau} - T_{\left( \frac{1}{2}, - \frac{1}{2} \right) \tau} \right) \psi$ & $1$ & $0$ \cr
\cr
$\frac{1}{\sqrt{2}} \left( T_{\left( \frac{1}{4},  - \frac{1}{4} \right) \tau} + T_{\left( \frac{3}{4}, - \frac{3}{4} \right) \tau} \right) \psi$ & $0$ & $1$ \cr
\cr
$\frac{1}{2} \left( T_{\left( \frac{l}{2 k^\prime}, -  \frac{l}{2 k^\prime} \right) \tau} {+} T_{\left( \frac{2 k^\prime - l}{2 k^\prime}, -  \frac{2 k^\prime - l}{2 k^\prime} \right) \tau} {+} T_{\left( \frac{k^\prime + l}{2 k^\prime}, -  \frac{k^\prime + l}{2 k^\prime} \right) \tau} {+}  T_{\left( \frac{k^\prime - l}{2 k^\prime}, -  \frac{k^\prime - l}{2 k^\prime} \right) \tau} \right) \psi$ & $l {+} 1 \mod 2$ & $1$ \cr
$\frac{1}{2} \left( T_{\left( \frac{l}{2 k^\prime}, -  \frac{l}{2 k^\prime} \right) \tau} {+} T_{\left( \frac{2 k^\prime - l}{2 k^\prime}, -  \frac{2 k^\prime - l}{2 k^\prime} \right) \tau} {-} T_{\left( \frac{k^\prime + l}{2 k^\prime}, -  \frac{k^\prime + l}{2 k^\prime} \right) \tau} {-}  T_{\left( \frac{k^\prime - l}{2 k^\prime}, -  \frac{k^\prime - l}{2 k^\prime} \right) \tau} \right) \psi$ & $l {+} 1 \mod 2$ & $0$ \cr
\cr \hline \cr
\end{tabular}
\caption{States in the $\SU (2) / \Z_2$ theory for $k^\prime = 2 \mod 4$ and trivial magnetic 't~Hooft flux $m_{12} = 0$. In the last two lines $1 \leq l \leq k^\prime / 2 - 1$.}
\label{k=2mod4table}
\end{table}
Here there are $(k^\prime - 2) / 4$ singlets with $(e_1, e_2) = (0, 0)$ and $(k^\prime + 2) / 4$ triplets with $(e_1, e_2) = (0, 1), (1, 0), (1, 1)$:
\bea
I^{\SO (3)}_{k^\prime} (0, 0; 0) & = & \frac{k^\prime - 2}{4} \cr
I^{\SO (3)}_{k^\prime} (e_1, e_2; 0) & = & \frac{k^\prime + 2}{4} \mathrm{\;\;\; for \;\;\;} (e_1, e_2) \neq (0, 0) 
\eea
so that
\bea \label{kprime=2mod4}
\hat{I}^{\SO (3)}_{k^\prime} (0, 0, 0) & = & k^\prime + 1 \cr
\hat{I}^{\SO (3)}_{k^\prime} (m_{23}, m_{31}, 0) & = & -1 \mathrm{\;\;\; for \;\;\;} (m_{23}, m_{31}) \neq (0, 0) .
\eea

We should also investigate the sector with non-trivial magnetic 't~Hooft flux $m_{12} = 1$. This means expanding around the unique flat connection which breaks $\SU (2) / \Z_2 \simeq \SO (3)$ completely yielding a unique state with trivial electric 't~Hooft flux $(e_1, e_2) = (0, 0)$. 

For $k^\prime = 0 \mod 4$ we must take this topologically non-trivial state to be bosonic so that
\bea
I^{\SO (3)}_{k^\prime} (0, 0; 1) & = & 1 \cr
I^{\SO (3)}_{k^\prime} (e_1, e_2; 1) & = & 0 \mathrm{\;\;\; for \;\;\;} (e_1, e_2) \neq (0, 0) 
\eea
or
\beq
\hat{I}^{\SO (3)}_{k^\prime} (m_{23}, m_{31}, 1) = 1 
\eeq
for all $m_{23}, m_{31} \in \Z / 2 \Z$. Together with (\ref{kprime=0mod4}), these results can be summarized  in an $\SL_2 (\Z)$ invariant way as
\bea
\hat{I}_{k^\prime} (0, 0, 0) & = & k^\prime + 1 \cr
\hat{I}_{k^\prime} (m_{23}, m_{31}, m_{12}) & = & 1 \mathrm{\;\;\; for \;\;\;} (m_{23}, m_{31}, m_{12}) \neq (0, 0, 0) . 
\eea

For $k^\prime = 2 \mod 4$ we must instead take the topologically non-trivial state to be fermionic so that
\bea
I^{\SO (3)}_{k^\prime} (0, 0; 1) & = & -1 \cr
I^{\SO (3)}_{k^\prime} (e_1, e_2; 1) & = & 0 \mathrm{\;\;\; for \;\;\;} (e_1, e_2) \neq (0, 0) 
\eea
i.e.
\beq
\hat{I}^{\SO (3)}_{k^\prime} (m_{23}, m_{31}, 1) =  -1 
\eeq
for all $m_{23}, m_{31} \in \Z / 2 \Z$. Together with (\ref{kprime=2mod4}) this means that
\bea
\hat{I}^{\SO (3)}_{k^\prime} (0, 0, 0) & = & k^\prime + 1 \cr
\hat{I}^{\SO (3)}_{k^\prime} (m_{23}, m_{31}, m_{12}) & = & -1 \mathrm{\;\;\; for \;\;\;} (m_{23}, m_{31}, m_{12}) \neq (0, 0, 0) . 
\eea

\subsection{$\hat{G} = \SU (3)$}
The elements of $\frac{1}{k^\prime} \Lambda^* / \Lambda$ can be represented by triples
\beq
(x_1, x_2, x_3) \in \left(\frac{1}{3 k^\prime} \Z / \Z \right)^3
\eeq
subject to the restrictions
\beq
x_1 + x_2 + x_3 = 0
\eeq
and 
\beq
x_1 = x_2 = x_3  \mod \frac{1}{k^\prime} \Z .
\eeq
We count these solutions modulo the action of the Weyl group (which permutes the three entries): Three solutions are given by the Weyl singlets
\beq
(x_1, x_2, x_3) = \left\{ \begin{array}{l} 
(0, 0, 0) \cr \cr
 \left( \frac{1}{3},  \frac{1}{3},  \frac{1}{3} \right) \cr \cr
 \left( \frac{2}{3},  \frac{2}{3},  \frac{2}{3} \right) 
\end{array} . \right.
\eeq
$3 k^\prime - 3$ further solutions are given by Weyl triplets of the form
\beq
(x_1, x_2, x_3) = (a, a, -2a), (a, -2 a, a), (-2 a, a, a)
\eeq
with $a \in \frac{1}{3 k^\prime} \Z / \Z$, $a \neq 0, \frac{1}{3}, \frac{2}{3}$. Finally, $\frac{1}{2} (k^\prime - 1) (k^\prime - 2)$ solutions are given by Weyl sextets 
\beq
(x_1, x_2, x_3), (x_2, x_3, x_1), (x_3, x_1, x_2), (x_3, x_2, x_1), (x_2, x_1, x_3), (x_1, x_3, x_2)
\eeq
with all three entries different.\footnote{
We may e.g. first choose $x_1$ in $3 k^\prime$ different ways and $x_2$ in $k^\prime$ differents ways, whereby $x_3$ is determined. But from these $3 (k^\prime)^2$ possibilities we must subtract $3 (3 k^\prime -3)$ for the Weyl triplets and $3$ for the Weyl singlets which have already been accounted for. The remaining $3 (k^\prime - 1) (k^\prime - 2)$ possibilities all fall into Weyl sextets.} Altogether we get 
\beq
I^{\SU (3)}_{k^\prime} = \frac{1}{2} (k^\prime + 2) (k^\prime + 1) 
\eeq
in agreement with (\ref{SU(n)groundstates}).

If $k^\prime = 3 s$ for some integer $s$, we may consider the theory with gauge group $\SU (3) / \Z_3$. The states described above then have trivial magnetic 't~Hooft flux $m_{12} = 0$. Among these, the distinguished Weyl invariant state 
\beq \label{distinguished}
\frac{1}{\sqrt{6}} \left( T_{\left(0, \frac{1}{3}, -\frac{1}{3} \right) \tau} + T_{\left(\frac{1}{3}, -\frac{1}{3}, 0 \right) \tau} + T_{\left(-\frac{1}{3}, 0, \frac{1}{3} \right) \tau} + T_{\left(0, -\frac{1}{3}, \frac{1}{3} \right) \tau} + T_{\left(-\frac{1}{3}, \frac{1}{3}, 0 \right) \tau} + T_{\left(\frac{1}{3}, 0, -\frac{1}{3} \right) \tau}  \right) \psi
\eeq
is invariant under the translations $T_{\left(\frac{1}{3}, \frac{1}{3}, -\frac{2}{3} \right) \tau}$ and $T_{\left(\frac{1}{3}, \frac{1}{3}, -\frac{2}{3} \right)}$ and thus has trivial electric 't~Hooft flux $(e_1, e_2) = (0, 0)$. The remaining Weyl invariant states fall into $\frac{1}{2} s (s + 1)$ multiplets with $9$ states each, one for each possible electric 't~Hooft flux $(e_1, e_2) \in (\Z / 3 \Z) \times (\Z / 3 \Z)$.\footnote{
Indeed, apart from the unordered triple $\left(0, \frac{1}{3}, -\frac{1}{3} \right)$ appearing in (\ref{distinguished}), all other triples transform non-trivially under translations by $\left(\frac{1}{3}, \frac{1}{3}, -\frac{2}{3} \right)$. This leads to an equal number of states for each value of $e_2 = 0, 1, 2$. By $\SL_2 (\Z)$ invariance, there must then be an equal number of states for each value of $e_1 = 0, 1, 2$.}
 The latter multiplet can be decomposed into an $\SL_2 (\Z)$ invariant state with $(e_1, e_2) = (0, 0)$ and an irreducible multiplet with one state for each $(e_1, e_2) \neq (0, 0)$. Altogether, we get
\bea
I^{\SU (3) / \Z_3}_{3 s} (0, 0; 0) & = & \frac{1}{2} s (s + 1) + 1 \cr
I^{\SU (3) / \Z_3}_{3 s} (e_1, e_2; 0) & = & \frac{1}{2} s (s + 1) \mathrm{\;\;\; for \;\;\;} (e_1, e_2) \neq (0, 0) ,
\eea
or after Fourier transformation
\bea
\hat{I}^{\SU (3) / \Z_3}_{3 s} (0, 0, 0) & = & \frac{9}{2} s (s + 1) + 1 \cr
\hat{I}^{\SU (3) / \Z_3}_{3 s} (m_{23}, m_{31}, 0) & = & 1 \mathrm{\;\;\; for \;\;\;} (m_{23}, m_{31}) \neq (0, 0) .
\eea
For each non-trivial magnetic 't~Hooft flux $m_{12} = 1, 2$, there is a single state, which has trivial electric 't~Hooft flux $(e_1, e_2) = (0, 0)$ and which must be taken to be bosonic. Thus
\bea
I^{\SU (3) / \Z_3}_{3 s} (0, 0; m_{12}) & = & 1 \cr
I^{\SU (3) / \Z_3}_{3 s} (e_1, e_2; m_{12}) & = & 0 \mathrm{\;\;\; for \;\;\;} (e_1, e_2) \neq (0, 0) ,
\eea
i.e.
\beq
\hat{I}^{\SU (3) / \Z_3}_{3 s} (m_{23}, m_{31}, m_{12}) = 1 
\eeq
for all $m_{23}, m_{31} \in \Z / 3 \Z$. Our results can be summarized in a manifestly $\SL_2 (\Z)$ invariant way as
\bea
\hat{I}^{\SU (3) / \Z_3}_{3 s} (0, 0, 0) & = & \frac{9}{2} s (s + 1) + 1 \cr
\hat{I}^{\SU (3) / \Z_3}_{3 s} (m_{23}, m_{31}, m_{12}) & = & 1 \mathrm{\;\;\; for \;\;\;} (m_{23}, m_{31}, m_{12}) \neq (0, 0, 0) .
\eea

The generalization of these results to $\hat{G} = \SU (n)$ for a prime number $n$ is straightforward. The cases where $n$ is not prime are more complicated because of the existence of proper subgroups of the center $C \simeq \Z / n \Z$. For the remaining simply laced groups, $\hat{G} = \Spin (4 k + 2)$ for which $C \simeq \Z / 4 \Z$ should be fairly straightforward, while $\hat{G} = \Spin (4 k)$ is slightly subtle because of its non-cyclic center $C \simeq (\Z / 2 \Z)^2$. The exceptional groups $E_6, E_7, E_8$ with $C \simeq \Z / 3 \Z, \Z / 2 \Z, 1$ might present a challenge because of their complicated Weyl groups.

\vspace*{3mm}
This research was supported by grants from the G\"oran Gustafsson foundation and the Swedish Research Council.

\end{document}